\input{aipcheck.tex}

\ifx\empty\selectedoptions
  \def\selectedoptions{final}
\fi

\documentclass[
  ]
  {aipproc}

\layoutstyle{8s}

\SetInternalRegister\hbadness{8000} 

\newcommand\doingARLO[2][]{%
  \ifx\mmref\undefined #1\else #2\fi
}

\def\etal{{\it et al.} }
\def\astpart{{\it Astrop. Physics}}

\def\apj{{\it ApJ}}
\def\apjl{{\it ApJ Lett.}}

\def\aas{{\it Astron.\ Astrophys.\ Suppl.}}

\def\inpress{{\it in press}}

\def\mic{{{\mu}m}}
\def\uk{{{\mu}K}}
\def\cm2{$cm^{-2}$}

\begin{document}

\title 
      []
      {Iterative map-making methods for Cosmic Microwave Background data analysis}


\author{Xavier Dupac}{
  address={Centre d'\'Etude Spatiale des Rayonnements,
9, avenue du Colonel Roche,
BP 4346,
F-31028 Toulouse cedex 4,
France,
dupac@cesr.fr}
}

\copyrightyear  {2001}

\begin{abstract}
The map-making process of Cosmic Microwave Background data involves
linear inversion problems which cannot be performed by a brute force approach
for the large timelines of most modern experiments.
We present optimal iterative map-making methods, both COBE and
Wiener, and apply these methods on simulated data.
They
produce very well restored maps, by removing nearly completely the correlated
noise that appears as intense stripes on the simply pixel-averaged maps.
\end{abstract}

\date{\today}

\maketitle

\section{Introduction}
The Cosmic Microwave Background (CMB hereafter) is being extensively studied nowadays,
thanks to the improvement of instrument and detector
performances.
Since the COBE experiment (Smoot \etal 1992), which performed the first detection of CMB
anisotropies, the size of time-ordered information (TOI hereafter) has been widely increased.
The data processing and analysis is thus still a
challenge for large timeline data, already in processing or still to
come.
The Cosmic Microwave Background data analysis is usually performed
in three steps, the first being the map-making process, the second the $C_{l}$
estimation from the maps and the noise covariance matrices, the third the
cosmological parameters estimation from the $C_{l}$ power spectrum.
In this article, we focus on the map-making step, which, in the context of large timelines, cannot be performed by a
brute-force approach, which would imply the manipulation and inversion of
tera-element large matrices.
Non optimal map-making methods have been developed, such as destriping using the scan
intercepts (see Delabrouille 1998 and Dupac \& Giard 2001).
We aim in this paper to apply the optimal methods to large timelines,
developing algorithms to avoid computation trouble.
The map-making methods for CMB are known to be optimal when following
desirable properties.
These are linear map-making methods, well known as the COBE method (Janssen
\& Gulkis 1992) and
the Wiener filter (Wiener 1949).
The matrix expressions of these optimal methods can be found, e.g., in Tegmark
(1997).

Applying these optimal methods to large timelines is not
straightforward, because of computer limitations.
The several tens or hundreds of mega-elements in a bolometer timeline
have to be processed by vector-only methods, that we aim to present in this
paper.
We mean by "vector-only methods" computations in which one never needs to
compute or even to store any large matrix.

\section{Observation strategy}
We present here a large-coverage observation strategy,
simulated from simple and usual technical requirements.
We simulate balloon-borne CMB experiments, required to scan the sky making
constant elevation circles at a rather constant rotation speed (as do
Archeops: Beno\^\i t \etal 2001, and TopHat: http://topweb.gsfc.nasa.gov).
This allows to observe a large area on the sky, thanks to the rotation of the
Earth (and, eventually, to the moving of the balloon on the Earth).
We have simulated a balloon-borne experiment timeline with a constant
scanning elevation of 35 degrees above the horizon, a sampling frequency of
100 Hz and a rotation speed of the gondola of 2 rpm.
The flight is 24 hours long, launched from a polar place (Kiruna for instance).
The winter time in this region provides polar nights allowing to make 24 hours
flight avoiding contamination from the sun.
The coverage for this polar flight is 35 \% of the sky.

\section{Simulation process}
We have made our simulated skies from three components: the CMB
fluctuations, the dipole, and the Galaxy, at 2 mm wavelength.
The simulated Universe we have simulated is $\Lambda$ dominated, with
$\Omega_\Lambda = 0.7$, $\Omega_{CDM} = 0.25$, $\Omega_{bar} = 0.05$, $H_0 = 50 $
and a scalar spectral index of the fluctuations equal to 1.
The $C_l$ simulated spectrum (with no cosmic variance) is made thanks
to the CMBFAST software (Seljak \& Zaldarriaga 1996).
The sky gaussian random field (i.e. a realization of a Universe with
the cosmological parameters we chose) is then simulated with the
SYNFAST tool of the HEALPix package (http://www.eso.org/science/healpix).
The dipole is added thanks to its COBE/DMR determination
(Lineweaver \etal 1996), and the Galaxy at 2 mm wavelength is extrapolated from
the composite 100 $\mic$ IRAS-COBE/DIRBE all sky dataset (Schlegel \etal 1998).
The noise we introduce in our simulated timelines can be characterized by its
statistical power spectrum which follows a 1/f law: $l_{inf} . (1 +
(f_c/f)^n)$, where $l_{inf}$ is the level of white noise (i.e. the only noise
at high frequency), $f_c$ the cut frequency and n the power index.
The noise we introduced is characterized by $n = 1$, $f_c = 0.1$ Hz and $l_{inf} = 100$ $\uk_{CMB}$ rms.
This white noise rms is approximatively the level expected for Planck
(Tauber 2000) bolometers with a 100 Hz sampling rate.
Of course this
quite low value is the level of the non-correlated noise in the
timelines,
but the total amount of noise introduced is much larger.

\section{Map-making methods applied on large timelines}

We use the HEALPix pixel scheme (http://www.eso.org/science/healpix) to make
our maps.
Reprojecting timelines on maps is not only a domain to domain
transform, but the simplest way to estimate the true map of the sky,
by averaging the samples of a same pixel on the sky. The noise is
therefore reduced by a factor square root of the number of samples in the
pixel (the weight).
We will not investigate the beam deconvolution in this article, and therefore
consider only 1-and-0 point-spread matrices.

The optimal map-making methods use the noise (N) and sky (S) covariance matrices:
these are impossible to invert or even to store for such large
timelines.
However, if the noise is stationary in the time domain, as it is the case for 1/f
noise and white noise (usual bolometer noises), the noise correlation matrix
in the time domain is circulant, i.e. multiplying a vector by this matrix is a
convolution, which is
filtering in the Fourier domain.
The sky covariance matrix is stationary in the map domain, as far as the Cosmic
Microwave Background is a gaussian random field (for timelines without the Galaxy).
In this case the sky covariance matrix in the map domain is circulant.

The COBE equation: \~x = $[A^{t} N^{-1} A]^{-1} A^{t} N^{-1} y$, cannot be directly applied with vector-only algorithms,
because of the matrix inversions needed.
(A is the point-spread matrix, N the noise covariance amtrix in the time
domain, y the data timeline and \~x the optimal reconstructed sky map.)
Thus the trick is to solve rather:

$[A^{t} N^{-1} A]$ \~x = $A^{t} N^{-1}$ y

This form prevents from the heavy inversion, but needs an iterative scheme.
The general iterative scheme for this equation is:

$\alpha$ \~x$_{n+1}$ = $\alpha$ \~x$_n$ + $A^{t} N^{-1}$ y - $[A^{t} N^{-1} A]$ \~x$_n$

where $\alpha$ is any linear operator on a vector, that is, any square matrix.
We have tested this algorithm on simulations and real data from the Archeops
experiment (Beno\^\i t \etal 2001), with $\alpha$ being a scalar.
By testing the method with different $\alpha$, we find that the identity is
the best iterator.

Another scheme can be developed, by making the noise map converge instead of
the sky map, as mentioned by Prunet (2001).
This can be better, as the signal can be more tricky than the instrumental
noise for the stability of the iterative scheme: hot galactic points for example
may induce stripes on the maps.
The noise-iterating scheme works with the following trick:
we change the variable \~x to \v{x} = $[A^tA]^{-1} A^t$ y - \~x.
It is straightforward to show that this is the noise map plus the
reconstruction error.
It leads to:

$\alpha$ \v{x}$_{n+1}$ = $\alpha$ \v{x}$_n$ + $A^{t} N^{-1}$ z - $[A^{t} N^{-1} A]$ \v{x}$_n$

where z = $A[A^tA]^{-1}A^t$ y - y.
If this algorithm converges, then the converging limit is exactly the optimal
solution of the map-making problem.
The case of $\alpha$ = I is actually the simplest iterator one can imagine,
but works well on the simulations and real data that we have processed.

The iterative scheme that we have developed for the Wiener method is close to the COBE one:

$\alpha$ \v{x}$_{n+1}$ = $\alpha$ \v{x}$_n$ + u - $[S^{-1} + A^{t} N^{-1} A]$
\v{x}$_n$

where u = $A^tN^{-1}[A[A^tA]^{-1}A^ty - y] + S^{-1}[A^tA]^{-1}A^ty$.
As we have shown for the COBE iterative method, here the converging limit is the
exact solution of the Wiener map-making equation.
This iterative scheme needs to handle both the N matrix, noise covariance
matrix in the timeline, that we process as a filter in the Fourier domain
like we do for the COBE iterative method, and the S matrix, sky covariance
matrix in the map domain.
Handling this as a matrix is not possible for a small scale pixelization that
we need for CMB experiments of today, thus we have to process it as a filter
in Fourier space, like we do for filtering timelines.
The HEALPix RING scheme is stationary with respect to the sphere, because it
pixelizes it making a ring around the sphere from the north pole to the south
pole, with equal pixel surfaces.
So filtering a HEALPix vector (i.e. a map) in Fourier space is optimal, to the
condition that there must not be large holes in the map, that would harm the
stationarity of the sky in the HEALPix scheme.

\section{Results and conclusion}

We have applied these methods to the polar flight simulated data:
the method reaches the residual noise at about 50 iterations, and this
residual is about 21.6 $\uk_{CMB}$ rms.
We can check that we have reached the convergence by observing the evolution
of the global residual noise rms, but also the evolution for some individual pixels, the map
aspect and the $C_l$ power spectrum.
We have to compare this result to the white noise amount in the map: the rms level of white noise in the polar flight map is 21.06 $\uk_{CMB}$ rms,
which is very close to the residual noise amount.
This shows how good the reconstruction is, as it is clear that the correlated
noise is significantly removed from the map.
The reconstructed map exhibits no visible difference with the true map.
The noise spectrum is nearly the one of a white noise above about l=50, but
exhibits some weak residual correlation at lower scales.
Even if the residual noise amount is very low, this deviation from the flat
spectrum could have to be taken into account for very precise measurements of
the low l.

This kind of vector-only methods seems to us unavoidable to make optimal maps
from CMB experiments of today, or still to come.
The reduction of the information in CMB data is a heavy work, from gigabytes of
rawdata to essentially 12 cosmological numbers with their error bars.
Since the computer facilities are limited and unsufficient for brute force
approaches (and it will be still the case for Planck data reduction), it is
an interesting challenge to process each step of this reduction work without
losing information.
Using stationarity properties of a signal in a given domain (sphere, map,
timeline...) to transform a matrix inversion problem into a vector-only
solution, could be probably also developed for other CMB reduction steps, such
as the component separation.

\section{Acknowledgements}
We would like to thank K.M. G\'orski and his collaborators for their so
useful HEALPix package.

\end{document}